\documentclass[journal,10pt]{IEEEtran}

\usepackage{ifpdf}
\usepackage{algorithmic}
\usepackage{algorithm}

\ifCLASSINFOpdf
\usepackage[pdftex]{graphicx}

\else
 \usepackage[dvips]{graphicx}
\fi
\usepackage[cmex10]{amsmath}
\usepackage{epstopdf}
\usepackage{array}
\usepackage{flushend}
\usepackage{balance}
\usepackage{eqparbox}

\usepackage[tight,footnotesize]{subfigure}
\usepackage{fixltx2e}
\usepackage{color}
\usepackage{float}

\usepackage{dblfloatfix}
\usepackage{url}
\usepackage{cite}
\usepackage{subfigure}
\usepackage{amsmath}
\usepackage{epstopdf}

\usepackage{amssymb}
\usepackage{color}

\begin{document}
\allowdisplaybreaks[3]
\title{Toward Massive MIMO HAPS Networks: Antenna Architectures and Design}
\setlength{\columnsep}{0.21 in}

\author{Shasha~Liu,~\IEEEmembership{Student Member,~IEEE,}
Abla~Kammoun,~\IEEEmembership{Member,~IEEE,}      
and
Mohamed-Slim~Alouini,~\IEEEmembership{Fellow,~IEEE}

\thanks {
Shasha Liu, Abla~Kammoun, and Mohamed-Slim Alouini are with the Division of Computer, Electrical and Mathematical Sciences and Engineering, King Abdullah University of Science and Technology, Thuwal 23955-6900, Saudi Arabia (e-mail: shasha.liu@kaust.edu.sa, abla.kammoun@kaust.edu.sa, slim.alouini@kaust.edu.sa).

}
}
\maketitle
\begin{abstract}
High-altitude platform stations (HAPS) are expected to become an important component of sixth-generation (6G) and beyond wireless networks by providing wide-area wireless connectivity through massive multiple-input multiple-output (mMIMO) technology. As the communication performance of HAPS systems is fundamentally determined by the antenna array architecture, efficient antenna design has become a critical research topic. Motivated by these considerations, this article provides a comprehensive overview of antenna array design for mMIMO-enabled HAPS systems and proposes two novel antenna architectures to improve antenna–user alignment and communication performance. Specifically, the representative applications of mMIMO-enabled HAPS are first introduced, followed by the key design requirements for HAPS antenna arrays. Representative antenna architectures are then systematically reviewed and compared under a unified simulation framework based on the 3GPP NTN channel model. 
Finally, several promising research directions are discussed to provide insights into the design of next-generation intelligent HAPS antenna systems.
\end{abstract}
\IEEEpeerreviewmaketitle
\section{Introduction}
Vertical heterogeneous networks (VHetNets) integrate satellite, aerial, and terrestrial networks for sixth-generation (6G) wireless communications, aligning with 3GPP's activities on non-terrestrial networks (NTNs) described in TR 38.811 \cite{3gpp_tr38811}. High-altitude platform stations (HAPSs) are a key component of the aerial layer.
Specifically, HAPSs operate in the stratosphere at altitudes of 17--22 km \cite{kurt2021vision}. Their quasi-stationary operation and favorable line-of-sight (LoS) propagation conditions enable wide-area air-to-ground connectivity.
Beyond direct access for ground users, HAPSs can provide backhaul connectivity for terrestrial and aerial base stations, facilitating terrestrial--NTN integration \cite{svistunov2025bridging}.
Compared with satellite systems, HAPSs offer lower propagation loss, reduced communication latency, and greater deployment flexibility. Compared with conventional low-altitude unmanned aerial vehicle (UAV)-mounted base stations, HAPSs can provide wider coverage, longer operational endurance, and greater payload capacity, depending on the platform design, enabling the deployment of large-scale antenna arrays. These characteristics make HAPSs an effective complement to terrestrial networks, particularly in rural, mountainous, and oceanic regions where communication infrastructure is limited.
\par
To satisfy the high-capacity communication requirements of HAPS while compensating for the considerable free-space path loss, massive multiple-input multiple-output (mMIMO) has emerged as a key enabling technology. By exploiting array gain, spatial multiplexing, and interference suppression, mMIMO significantly improves spectral efficiency, network capacity, and communication reliability. However, realizing these gains critically depends on the underlying antenna array architecture. In HAPS systems, the array geometry and antenna element orientations jointly determine the radiation characteristics toward different ground locations, thereby affecting coverage, beamforming capability, and spatial multiplexing performance. 
In particular, three-dimensional antenna architectures exploit flexible element placement to improve the spatial separation of user channels, potentially reducing inter-user channel correlation \cite{shafie2026novel}. Their diverse element orientations also allow different parts of the array to serve users in different azimuth and elevation directions, extending coverage \cite{abbasi2024hemispherical}.


Various antenna array architectures have been proposed for HAPS, ranging from conventional planar arrays \cite{el2001cellular} to cylindrical \cite{tashiro2021cylindrical}, hexagonal \cite{3gpp_tr38901}, sectorized polygonal \cite{shafie2026novel}, and hemispherical arrays \cite{abbasi2024hemispherical}.
Conventional planar arrays provide high antenna gain for users directly beneath the HAPS but suffer from limited coverage at the cell edge. Cylindrical arrays extend the coverage area by deploying antenna elements on both the side surface and the bottom panel, whereas hexagonal arrays improve the radiation toward edge users through tilted side panels. 
Similarly, a sectorized architecture uses multiple vertical planar antenna panels arranged to form a polygonal prism, with each panel serving a distinct azimuth sector \cite{shafie2026novel}.
Hemispherical array structures have been proposed to enhance angular coverage by better aligning antenna elements with users over a wide range of directions. Although numerous HAPS antenna array architectures have been proposed, they are typically developed and evaluated independently. Consequently, the underlying design ideas and the resulting communication performance of different antenna architectures have not been systematically analyzed and compared.
\par
Motivated by these observations, this article provides a comprehensive review of HAPS antenna array architectures from the perspective of mMIMO communications. We first introduce the representative applications of mMIMO-enabled HAPS networks and identify the key design requirements for HAPS antenna arrays. Representative antenna architectures are then systematically reviewed and evaluated under a unified simulation framework based on the 3GPP NTN channel model \cite{3gpp_tr38811}, enabling a fair comparison of their communication performance under diverse user distributions and coverage scenarios. In addition, two novel antenna architectures, namely the truncated cone array and the truncated hemispherical array, are proposed to improve antenna-user alignment and communication performance. Finally, several promising research directions, including movable antenna-enabled arrays, electromagnetic–communication co-design, metamaterial-enabled antenna arrays, non-uniform array architectures, and AI-driven antenna optimization, are discussed to provide insights into the development of next-generation intelligent HAPS antenna systems.

\section{Massive MIMO for HAPS Networks}
mMIMO has become a key enabling technology for modern wireless communication systems. Owing to their large payload capacity and physical size, HAPS platforms can accommodate large-scale antenna arrays, enabling highly directional beamforming, substantial array gains, and spatial multiplexing to significantly enhance network coverage and capacity. The following discusses the key advantages and representative applications of mMIMO in HAPS networks.
\subsection{HAPS as an Aerial Super Macro Base Station}
Owing to their quasi-stationary operation, wide-area coverage, and favorable LoS propagation, HAPS platforms can operate as aerial super macro base stations (HAPS-SMBSs), providing seamless broadband access over extensive geographical areas. Compared with conventional terrestrial macro base stations, HAPS-SMBSs can simultaneously serve a significantly larger number of users while extending wireless connectivity to underserved and infrastructure-limited regions. Beyond conventional wireless access, HAPS-SMBSs can further evolve into intelligent aerial platforms by integrating communication, computation, and caching capabilities, thereby supporting emerging 6G and beyond services such as edge intelligence, real-time data processing, and content delivery.

\subsection{Supporting Temporary Traffic Hotspots}
Temporary and unpredictable traffic hotspots, such as large public gatherings, sporting events, and emergency situations, can easily overload terrestrial communication networks. Owing to their deployment flexibility, HAPS can complement terrestrial infrastructure by providing additional wireless capacity where it is needed. Furthermore, by exploiting mMIMO, HAPS can simultaneously serve a large number of users through highly directional beamforming and spatial multiplexing. This capability is particularly effective in HAPS systems because the air-to-ground channel is dominated by strong LoS propagation and a small angular spread, as characterized in the 3GPP channel model \cite{3gpp_tr38811}. These favorable propagation characteristics enable mMIMO to achieve high beamforming gains and efficient spatial multiplexing, thereby substantially improving network capacity during temporary traffic surges.

\subsection{Supporting Future Intelligent Transportation Networks}
The emergence of intelligent transportation systems (ITSs) is extending beyond ground vehicles to include flying cars and electric vertical takeoff and landing (eVTOL) aircraft, imposing stringent requirements on seamless three-dimensional connectivity \cite{kurt2021vision}. Operating at an altitude of 17–22 km, HAPS is naturally positioned to provide wireless services for both terrestrial and low-altitude aerial users. In particular, HAPS can complement terrestrial networks by extending coverage to areas where ground base stations are unavailable or blocked, such as mountainous regions, rural areas, and urban canyons, thereby enabling continuous connectivity for future ITS applications. Furthermore, mMIMO significantly enhances the communication capacity of HAPS while maintaining robust beamforming performance in high-mobility environments, making it well-suited for supporting flying cars and other next-generation intelligent transportation services.


\subsection{Supporting Coverage Gap Filling}
HAPS can effectively complement terrestrial networks by providing wireless connectivity in coverage gaps caused by terrain blockage, dense urban environments, or insufficient signal strength from terrestrial base stations (BSs). Such coverage deficiencies become even more pronounced in millimeter-wave (mmWave) networks due to their high susceptibility to blockage. Benefiting from their large platform size and quasi-stationary deployment, HAPS-SMBSs can employ large-scale antenna arrays to generate highly directional three-dimensional (3D) beams toward users in blocked areas. This capability enables HAPS to efficiently restore coverage and maintain reliable connectivity, particularly in challenging propagation environments.

\section{Design Requirements for HAPS Antenna Array}

The diverse HAPS applications discussed in Section~II impose
different requirements on antenna array design, depending on
the spatial distribution and mobility of the served users or nodes.
Table~\ref{tab:application_requirements} summarizes these connections
by mapping representative applications to their typical user or
node distributions and key antenna design requirements.
Building on these connections,
the following subsections discuss three key aspects of HAPS
antenna array design: wide-area coverage, spatial multiplexing
and beamforming capability, and practical implementation feasibility.

\begin{table*}[!h]
\centering
\caption{Representative HAPS applications, user distributions, and antenna design requirements.}
\label{tab:application_requirements}
\renewcommand{\arraystretch}{1.3}
\begin{tabular}{|p{3.0cm}|p{4.5cm}|p{7.5cm}|}
\hline
\textbf{Application} &
\textbf{Typical User/Node Distribution} &
\textbf{Key Antenna Design Requirements} \\
\hline

Aerial super macro base station &
Users dispersed over a wide area, with possible local concentrations. &
Broad azimuth and elevation coverage; boresight distribution
balancing wide-area gain and local demand; multiuser spatial multiplexing. \\
\hline

Temporary traffic hotspots &
Dense user clusters within localized regions. &
Concentrated gain toward hotspots; sufficient angular resolution
and channel separability for interference suppression. \\
\hline

Intelligent transportation &
Mobile ground vehicles and low-altitude aerial users. &
Diverse antenna element orientations for broad azimuth and
elevation coverage; beam steering and tracking for mobile users. \\

\hline

Coverage gap filling &
Users in localized, irregularly distributed underserved regions. &
High antenna gain toward underserved regions;
multiple beams to serve geographically separated coverage gaps.\\
\hline
\end{tabular}
\end{table*}

\subsection{Wide-Area Coverage}
One of the primary objectives of HAPS systems is to provide seamless wireless connectivity over a wide geographical area. This requires antenna designs that account for both elevation-angle coverage and element-boresight distribution. For a ground user at a horizontal distance $r$ from a HAPS at altitude $H$, the off-nadir angle is $\theta(r)=\tan^{-1}(r/H)$. With $H=20$ km and a coverage radius of $100$ km, this angle ranges from $0^\circ$ at the nadir to approximately $78.7^\circ$ at the cell edge. The array should therefore provide sufficient gain over a broad range of elevation directions.
In addition, within this angular range, the distribution of element boresights also determines how antenna gain is allocated across the service area. Since an element's gain decreases as the user direction deviates from its boresight, diverse orientations in both azimuth and elevation allow different subsets of elements to provide favorable gain toward users at different locations. Moreover, the number of elements oriented toward each region influences the gain available to users there. Antenna design should therefore consider both the range of boresight directions and their distribution, balancing wide-area coverage with enhanced gain toward regions of higher user concentration.

\subsection{Spatial Multiplexing and Beamforming Capability}

Beyond wide-area coverage, HAPS requires sufficient spatial degrees
of freedom to serve multiple users simultaneously. Intra-user spatial
correlation and inter-user channel similarity can limit spatial
multiplexing and interference suppression, respectively. Antenna element positions and orientations should therefore be jointly designed to address both aspects while maintaining favorable antenna–user alignment.
Specifically, 
for an individual user, element orientations and positions jointly affect spatial correlation. Differently oriented elements emphasize multipath components
arriving from different directions, which can reduce the
correlation between their received signals, while three-dimensional element placement can introduce direction-dependent phase differences that partially cancel when averaged over the multipath arrival directions. Together, these effects can help reduce spatial correlation and improve the effective channel dimensionality.
For different users, channel separability depends on the differences in their gain and phase profiles across the array. In a planar array, the common element orientation gives each user approximately the same element gain across the array under the far-field assumption. User separation therefore relies primarily on geometry-dependent phase differences. Three-dimensional arrays with diverse element orientations introduce additional gain-profile differences between users at different azimuth and elevation angles. Together with the phase differences determined by element positions, these gain variations can improve inter-user channel separability and facilitate interference suppression.


\subsection{Practical Implementation Feasibility}
Besides communication performance, practical HAPS antenna design must account for hardware and platform constraints. 
Stable HAPS operation relies on flight control, energy management, and communication subsystems that share limited payload capacity and energy resources, constraining the weight and power consumption of onboard communication equipment \cite{kurt2021vision,svistunov2025bridging}.
Antenna architectures should therefore be evaluated in terms of payload mass, power consumption, mechanical complexity, aerodynamic compatibility, and implementation cost. 
Radio-frequency (RF) chain mapping and hybrid beamforming (HBF) feasibility are equally important. Fully digital beamforming typically requires a dedicated RF chain for each antenna element, which can impose substantial power and hardware demands on large-scale arrays. HBF can reduce the RF-chain count by combining analog processing across antenna elements with digital processing across RF chains. However, the required number of RF chains and the achievable performance depend on subarray connectivity, analog-network constraints, and spatial multiplexing requirements, rather than array geometry alone.

\section{HAPS Antenna Array Architectures and Design Evolution}

\begin{table*}[t]
\caption{Representative HAPS antenna array architectures,
their design philosophies, and simulation configurations.}
\label{tab:architecture_summary}
\renewcommand{\arraystretch}{1.4}
\begin{tabular}{|m{2.6cm}|m{4.0cm}|m{6.0cm}|m{4.0cm}|}
\hline
\textbf{Architecture} & \textbf{Configuration} & \textbf{Design Idea} & \textbf{Simulation Configuration} \\
\hline

Planar array &
\includegraphics[
    width=0.2\textwidth,
    trim=10 10 10 10,
    clip
]{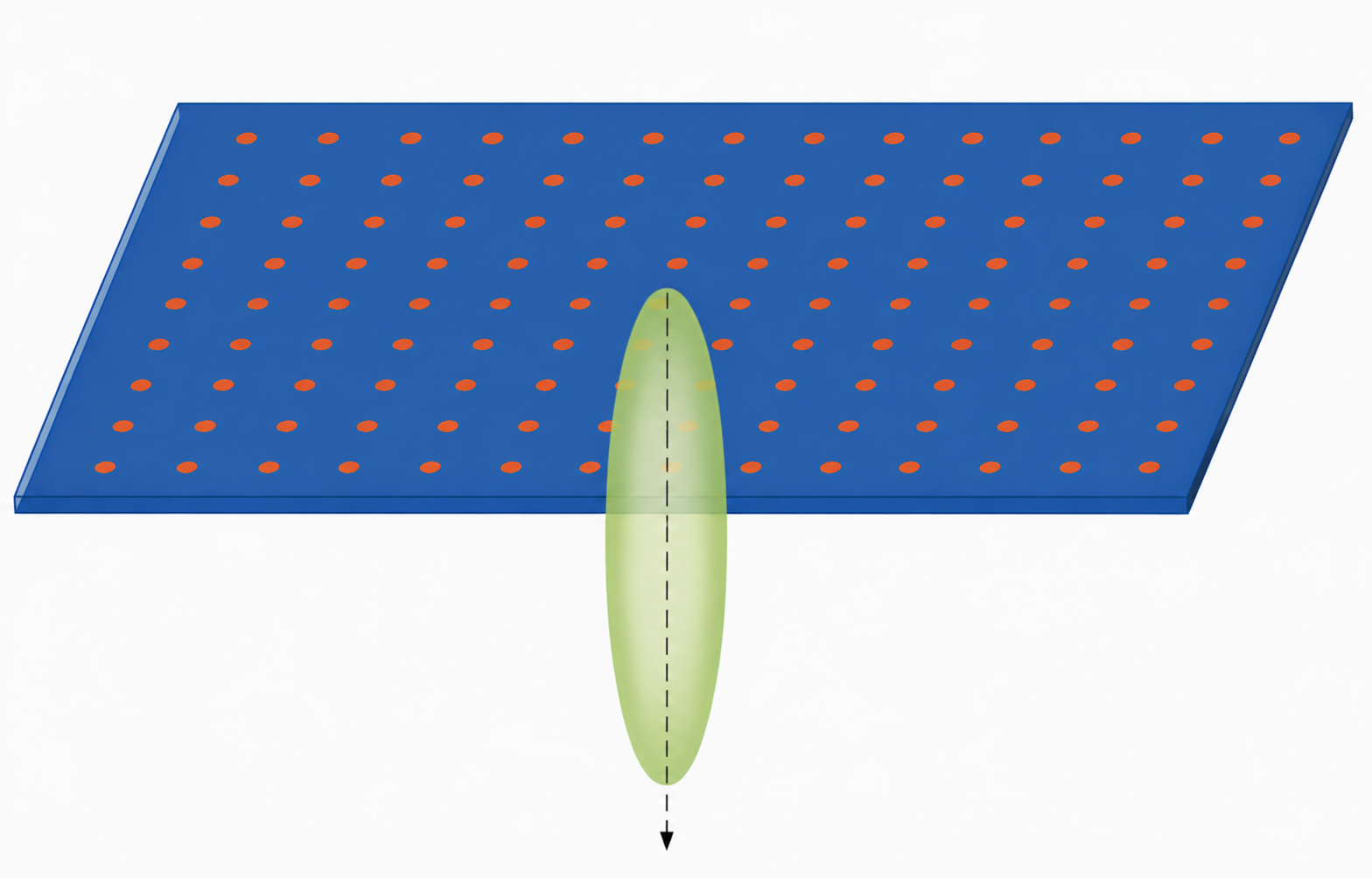} &
All antenna elements are mounted on a single downward-facing planar panel, providing a simple baseline architecture for HAPS communications.
& A $20 \times 10$ uniform planar array on the downward-facing panel.
\\
\hline

Cylindrical array &
\includegraphics[width=0.2\textwidth,
trim=10 10 10 10,
    clip
]{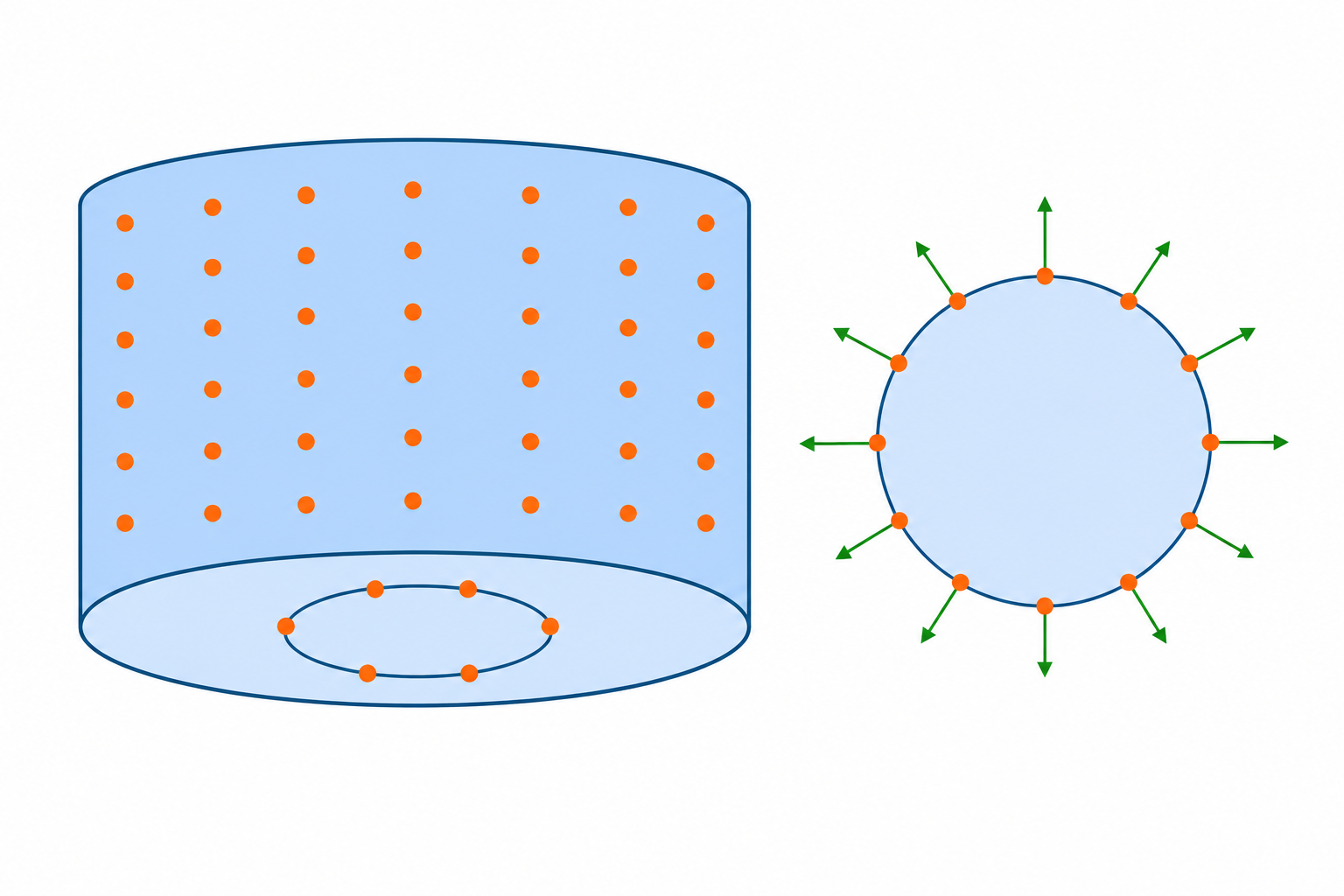} &
Antenna elements are deployed on the cylindrical side surface with boresights parallel to the ground, together with a bottom panel, to extend coverage from the nadir to the cell edge 
&
Side surface: 6 circular rings, each containing 32 elements.
\par Bottom surface: 8 uniformly spaced elements in a circular array.\\
\hline

3GPP Hexagonal array &
\includegraphics[width=0.2\textwidth,
trim=10 10 10 10,
    clip
]{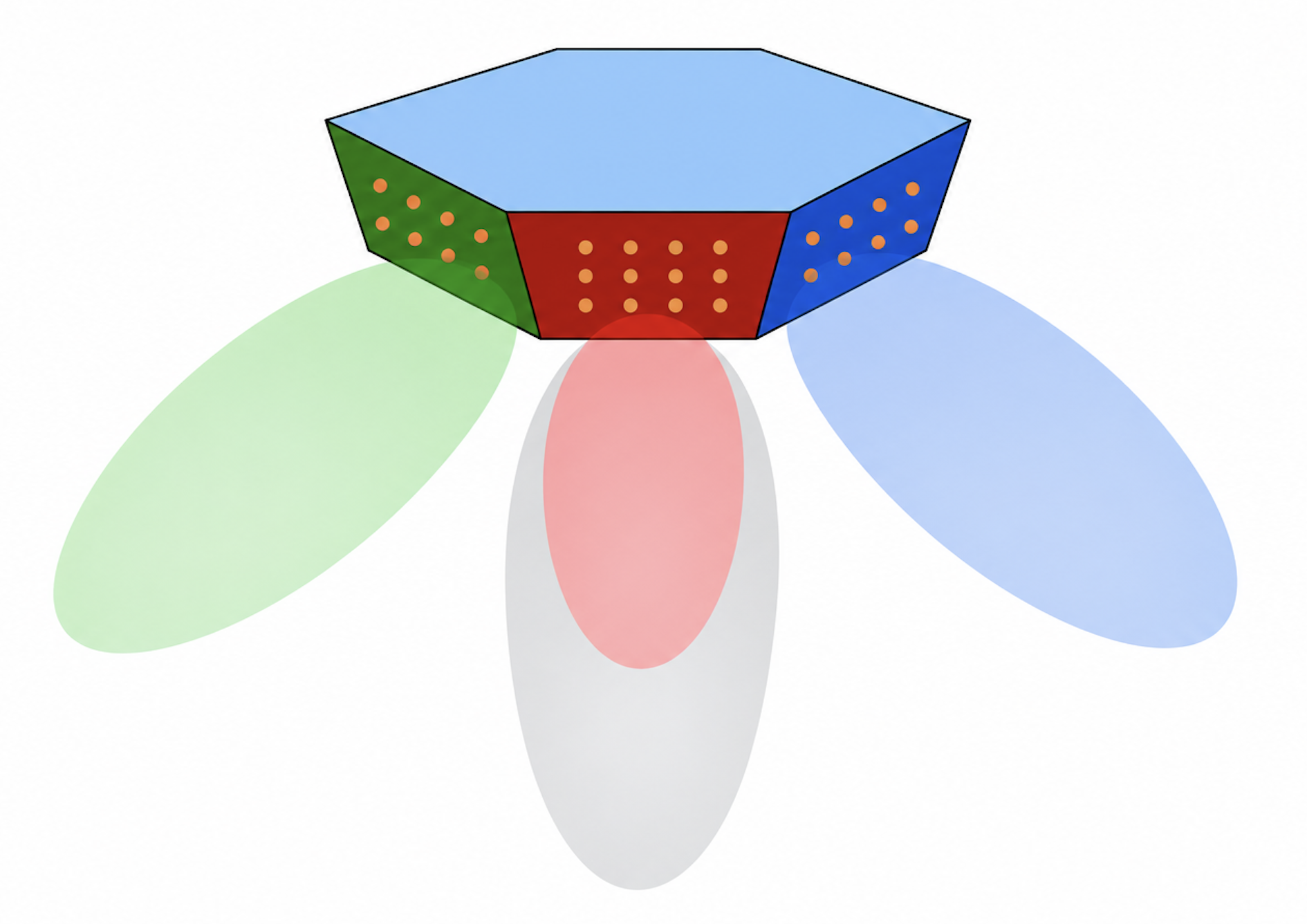} &
Six downward-tilted side panels and one bottom panel are employed to better align the antenna boresights with ground users, thereby improving the antenna gain over a wide coverage area. 
& Side panels: 6 panels, each containing a $4 \times 8$ array.
\par Bottom panel: a $2 \times 4$ array. \\
\hline

Truncated Cone array&
\includegraphics[width=0.2\textwidth]{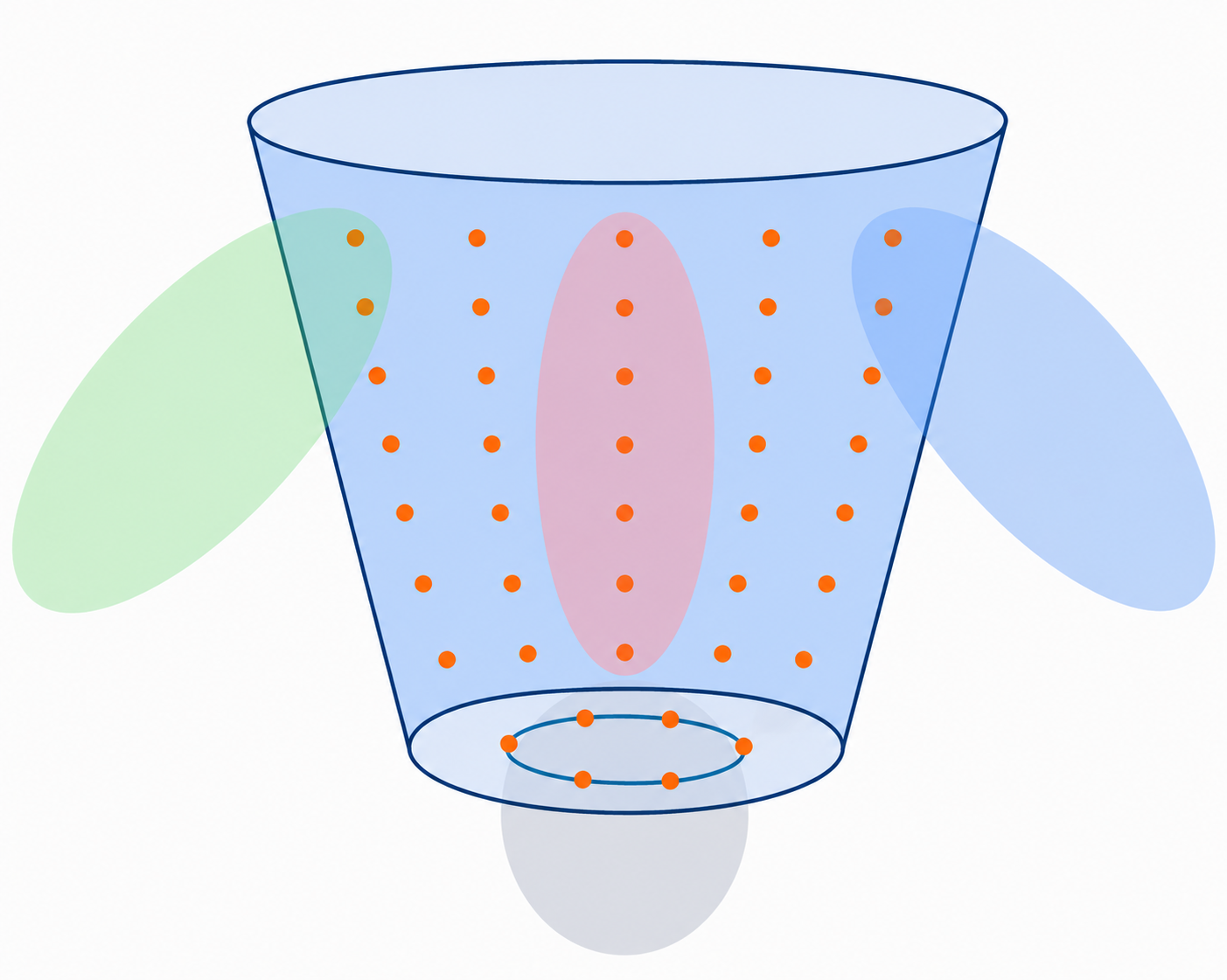} &
The cylindrical side surface is tilted toward the ground to improve antenna–user alignment while preserving continuous azimuthal coverage and the bottom panel for nadir coverage.
&
Side surface: 6 circular rings, each containing 32 elements.
\par Bottom surface: 8 uniformly spaced elements in a circular array\\
\hline

Hemispherical array&
\includegraphics[width=0.15\textwidth]{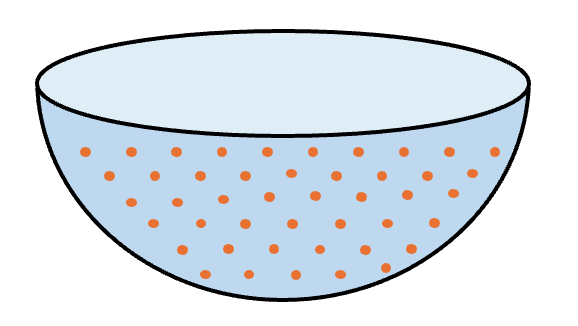} &
Antenna elements are distributed over a hemispherical surface to increase orientation flexibility and provide continuous three-dimensional angular coverage. 
& 6 circular rings containing 39, 38, 37, 34, 29, and 23 elements, respectively, for a total of 200 elements. 
\\
\hline
Truncated Hemispherical array&
\includegraphics[width=0.15\textwidth]{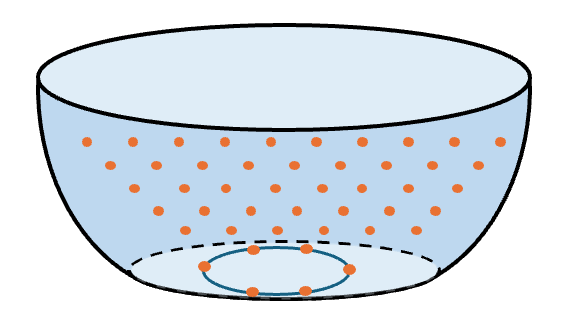} &
An additional bottom panel is introduced to enhance antenna gain for nadir users while preserving the continuous diversity of orientations of the hemispherical array.
&
Curved surface: 6 circular rings containing 38, 37, 35, 33, 28, and 21 elements, respectively.
\par Bottom surface: 8 uniformly spaced elements in a circular array. 
\\
\hline
\end{tabular}
\end{table*}

To address the design requirements discussed in the previous section,
various antenna array architectures have been proposed for HAPS
massive MIMO. These architectures differ in array geometry and
element orientation, leading to different coverage and beamforming
characteristics. Table~\ref{tab:architecture_summary} summarizes
representative architectures and their design rationales.
The following subsections discuss existing and proposed architectures
and compare their performance under a common simulation setup.

\par
To facilitate a consistent performance comparison, we consider an uplink single-input multiple-output (SIMO)
 system in which a single HAPS at an altitude of $H=20$ km serves $K$ non-cooperating single-antenna user equipments (UEs) distributed within a circular ground area of radius $R_{\mathrm{radius}}$ km, centered at the HAPS ground projection. Unless otherwise stated, $K=20$ UEs are served simultaneously.
 All array architectures employ $M=200$ antenna elements with a nominal inter-element spacing of $0.7\lambda$ and fully digital minimum mean-square error (MMSE) combining with one RF chain per element. Thus, the antenna-element and RF-chain counts are identical across architectures, while their physical apertures and dimensions may differ. In addition, the element arrangement and allocation across array surfaces are detailed in the last column of
Table~\ref{tab:architecture_summary}.
\par
Each antenna element follows the 3GPP directional radiation pattern~\cite{3gpp_ntn_rf_r18}, with a maximum gain of $G_{E,\max}=8$ dBi and half-power beamwidths of $65^\circ$ in both the azimuth and elevation planes. The element pattern is evaluated in the local coordinate system of each element to account for its orientation. The array response accounts for the three-dimensional element positions, and its detailed expressions for each architecture are provided in~\cite{liu2026uplink}.
\par
The UE--HAPS channel is modeled as a Rician channel comprising a LoS component and clustered non-line-of-sight (NLoS) components which are given in equation (22) in \cite{liu2026uplink}. In addition, the UE transmit power and antenna gain are $P_t=23$ dBm and $G_{\mathrm{UE}}=0$ dBi, respectively. The receiver noise power is $\sigma_{\mathrm{dBm}}^2=-174+10\log_{10}(\mathrm{BW})+\mathrm{NF}$, with bandwidth $\mathrm{BW}=1.08\times10^6$ Hz and noise figure $\mathrm{NF}=9$ dB.

\par
Performance is evaluated using the cumulative distribution function (CDF) of the sum rate:
$
F_{R_{\mathrm{sum}}}(r_{\mathrm{th}})
=
\Pr\left\{R_{\mathrm{sum}}\leq r_{\mathrm{th}}\right\},
$
where
$
R_{\mathrm{sum}}
=
\mathrm{BW}\sum_{k=1}^{K}
\log_2\left(1+\mathrm{SINR}_k\right)
$
is the sum rate in bits/s, and $\mathrm{SINR}_k$ is the post-combining signal-to-interference-plus-noise ratio of UE $k$. 

\par


\begin{figure}[!t]
\centering
\subfigure[Uniformly distributed UEs.]{%
    \includegraphics[width=3in]{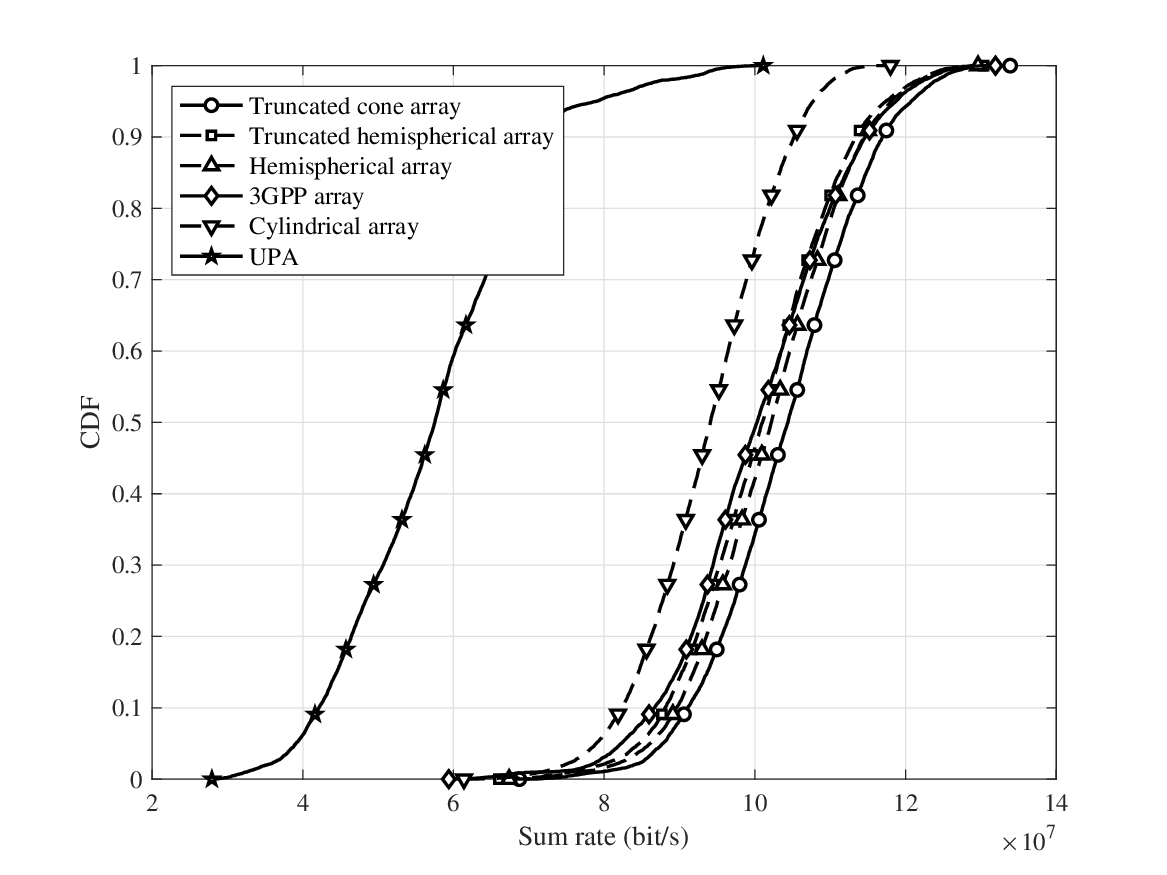}
    \label{Fig:uniform_K20}
}\\
\subfigure[Gaussian-distributed UEs with $\sigma=80$ km.]{%
    \includegraphics[width=3in]{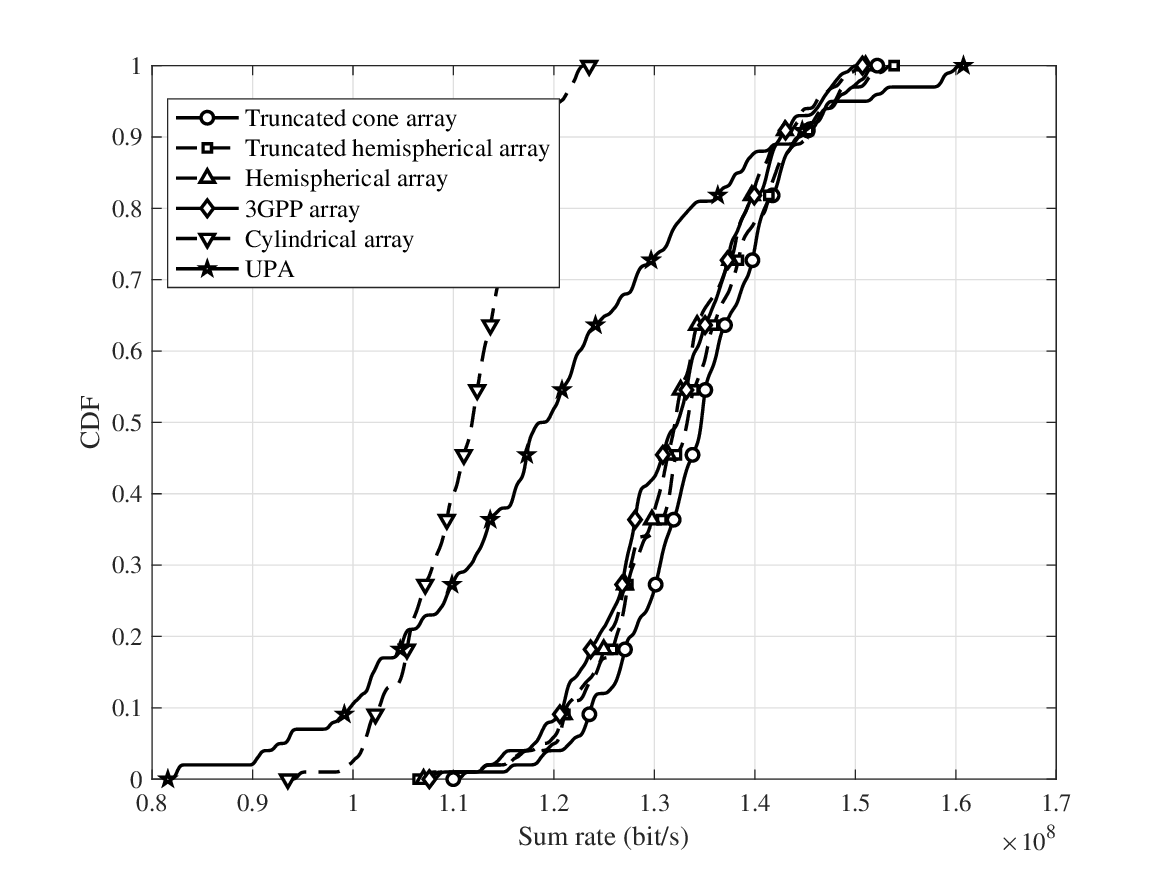}
    \label{Fig:Gaussian_sigma80}
}
\caption{Sum-rate CDFs for different UE distributions within
a circular cell of radius $R_{\mathrm{radius}}=100$ km.}
\label{Fig:sum_rate_R100}
\end{figure}

\subsection{Existing HAPS Antenna Array Architectures}
\subsubsection{Planar Array}
\begin{figure}[!t]
\centering
\subfigure[Uniformly distributed UEs with $R_{\mathrm{radius}}=20$ km.]{%
    \includegraphics[width=3in]{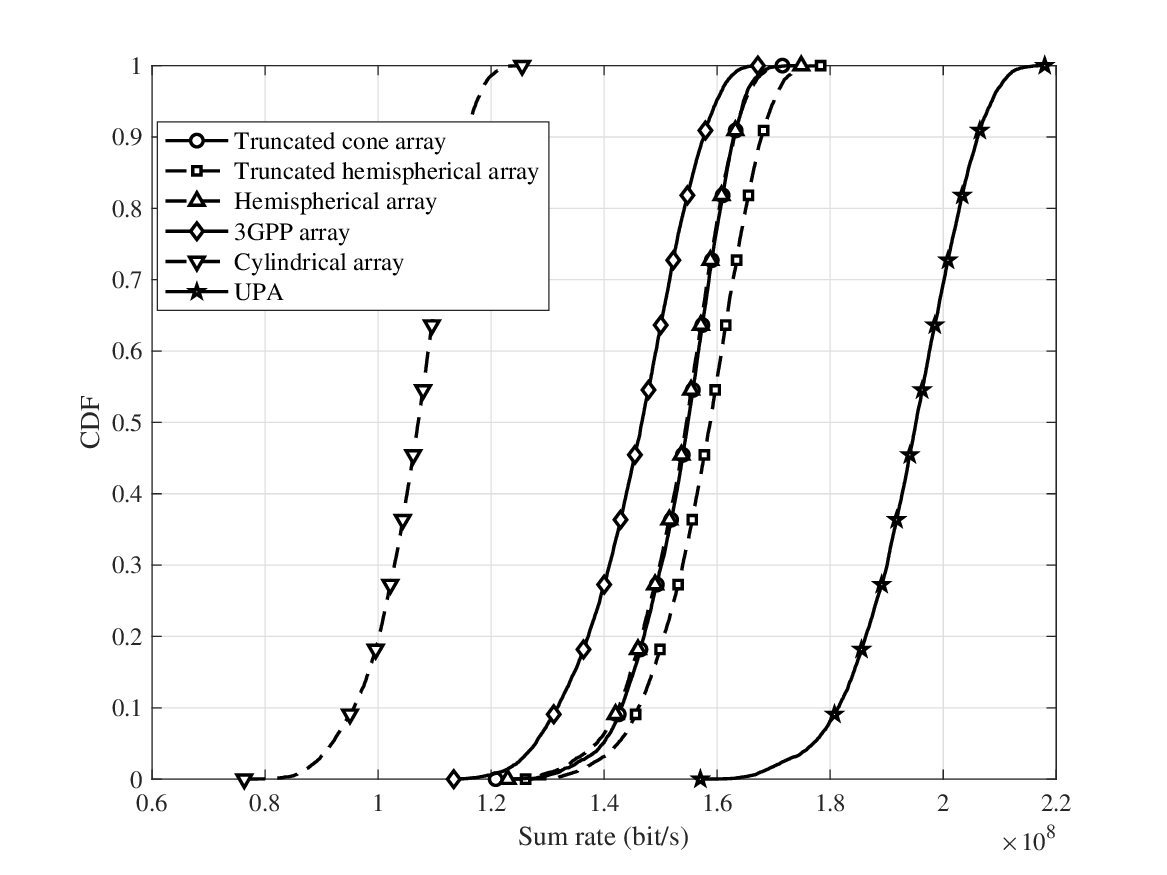}
    \label{Fig:uniform_K20_R20}
}\\
\subfigure[Gaussian-distributed UEs with $R_{\mathrm{radius}}=100$ km and $\sigma=10$ km.]{%
    \includegraphics[width=3in]{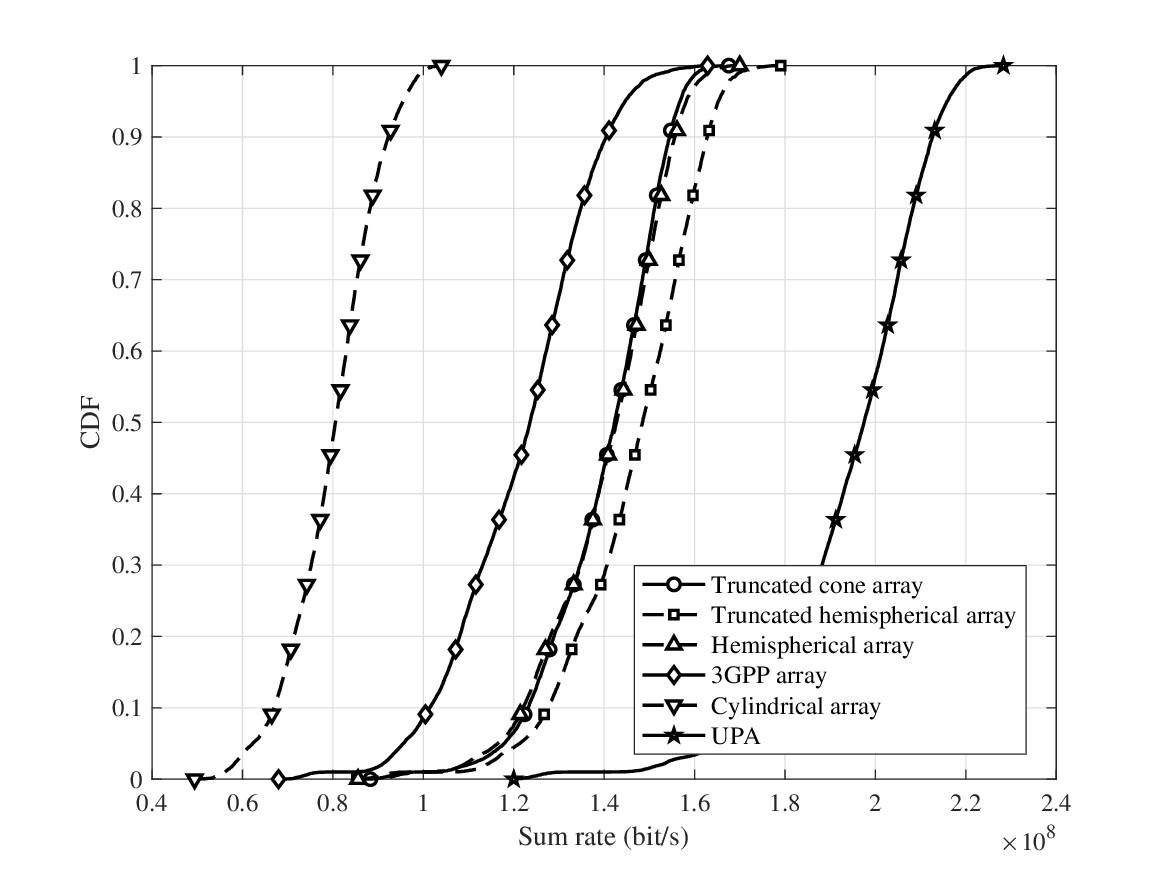}
    \label{Fig:Gaussain_sigma10}
}
\caption{Sum-rate CDFs for UEs concentrated near the HAPS nadir.}
\label{Fig:sum_rate_concentrated}
\end{figure}

Uniform planar antenna arrays (UPAs) have been proposed for HAPS in conventional systems \cite{el2001cellular}. As shown in the first row of Table~\ref{tab:architecture_summary}, the planar array consists of uniformly distributed antenna elements mounted on a flat panel, with all elements pointing vertically downward. While this simple architecture provides a convenient baseline for HAPS antenna design, its fixed boresight limits the antenna gain for users located away from the nadir direction. This limitation is evident in Fig.~\ref{Fig:sum_rate_R100}, which compares the
sum-rate CDFs for uniformly and Gaussian-distributed UEs
within a circular service area of radius
$R_{\mathrm{radius}}=100$ km. 
In both scenarios, the planar array generally achieves lower
sum rates than most of the considered architectures. Although users near the nadir benefit from high antenna gain, those farther from the nadir experience substantial gain degradation, limiting the aggregate throughput when users are widely dispersed. Consequently, planar arrays are mainly suitable for HAPS systems with relatively small coverage radii.
This observation is further validated by Fig.~\ref{Fig:sum_rate_concentrated}. Specifically, when the coverage radius is reduced to 20 km under the uniform UE distribution or when UEs follow a highly concentrated Gaussian distribution with $\sigma=10$ km, representing hotspot traffic scenarios, UPA achieves the best communication performance.
From an implementation perspective, the flat-panel geometry simplifies fabrication, mechanical mounting, and feeding-network layout, while its regular element arrangement facilitates subarray partitioning and RF-chain mapping for hybrid beamforming. These features make planar arrays attractive for small-coverage or hotspot deployments.


\subsubsection{Cylindrical Array}
To overcome the limited coverage of planar arrays, SoftBank proposed the cylindrical antenna array architecture for HAPS systems \cite{tashiro2021cylindrical}. As illustrated in the second row of Table~\ref{tab:architecture_summary}, antenna elements are deployed on both the cylindrical side surface and the bottom panel, allowing the array to simultaneously serve users near the nadir and those close to the coverage boundary. As shown in Fig.~\ref{Fig:uniform_K20}, the cylindrical array achieves a higher sum rate than the planar array. However, the side-mounted antenna elements are oriented parallel to the ground, whereas the directions toward most ground users have a downward elevation angle. Consequently, the boresights of these elements are not well aligned with the user directions, leading to reduced antenna gain, especially for users at intermediate and high elevation angles.
The cylindrical geometry increases support and feeding-network complexity compared with planar arrays, while its regular layout facilitates sector-based subarray partitioning for hybrid beamforming. Its wider coverage must therefore be weighed against these additional implementation requirements.

\subsubsection{3GPP Array}
To further improve coverage, 3GPP proposed a hexagonal antenna architecture, as shown in the third row of Table~\ref{tab:architecture_summary} \cite{3gpp_tr38901}. Compared with the planar array, the six side panels together with one bottom panel provide additional azimuth-angle freedom, enabling wider spatial coverage. Furthermore, unlike the cylindrical array, the tilted side panels better align the antenna boresights with ground users over a broad range of elevation angles. 
These advantages are reflected in Fig.~\ref{Fig:sum_rate_R100}, 
where the 3GPP array consistently achieves a higher sum rate than both the planar and cylindrical arrays, demonstrating the benefits of combining azimuthal orientation diversity with appropriately chosen panel tilt angles. The hexagonal architecture uses flat panels that facilitate modular fabrication and panel-based RF-chain mapping for hybrid beamforming.
The use of multiple tilted panels nevertheless increases mounting and inter-panel calibration complexity compared with a single planar panel.

\subsubsection{Hemispherical Array}
Different from cylindrical and panel-based arrays, the hemispherical array proposed in \cite{abbasi2024hemispherical} provides greater flexibility in antenna element orientations. As shown in the fifth row of Table~\ref{tab:architecture_summary}, antenna elements are uniformly distributed over a hemispherical surface, providing a continuous range of elevation angles. As illustrated in Fig.~\ref{Fig:sum_rate_R100}, the hemispherical array consistently outperforms the cylindrical and 3GPP arrays. This advantage is maintained for both a smaller coverage radius of 20 km (Fig.~\ref{Fig:uniform_K20_R20}) and a hotspot scenario with Gaussian-distributed UEs ($\sigma=10$ km, Fig.~\ref{Fig:Gaussain_sigma10}), demonstrating the robustness of the hemispherical architecture across diverse deployment scenarios. However, because relatively few antenna elements are oriented toward the nadir direction, the achievable antenna gain for users directly beneath the HAPS remains limited, which may degrade performance in hotspot scenarios where users are densely concentrated below the HAPS, such as large public gatherings or football stadiums.
The hemispherical surface requires complex three-dimensional fabrication and feeding networks, while element orientations varying in both azimuth and elevation complicate subarray partitioning, RF-chain mapping, and calibration for hybrid beamforming.

\subsection{Proposed HAPS Antenna Array Architectures}
Motivated by the limitations of existing architectures, we propose truncated cone and truncated hemispherical arrays for HAPS communications. These designs modify both element positions and boresight orientations. Their design principles, communication performance, and practical considerations are discussed below.
\subsubsection{Truncated Cone Array}
\begin{figure}[h!]
\centering
\includegraphics[width=3in]{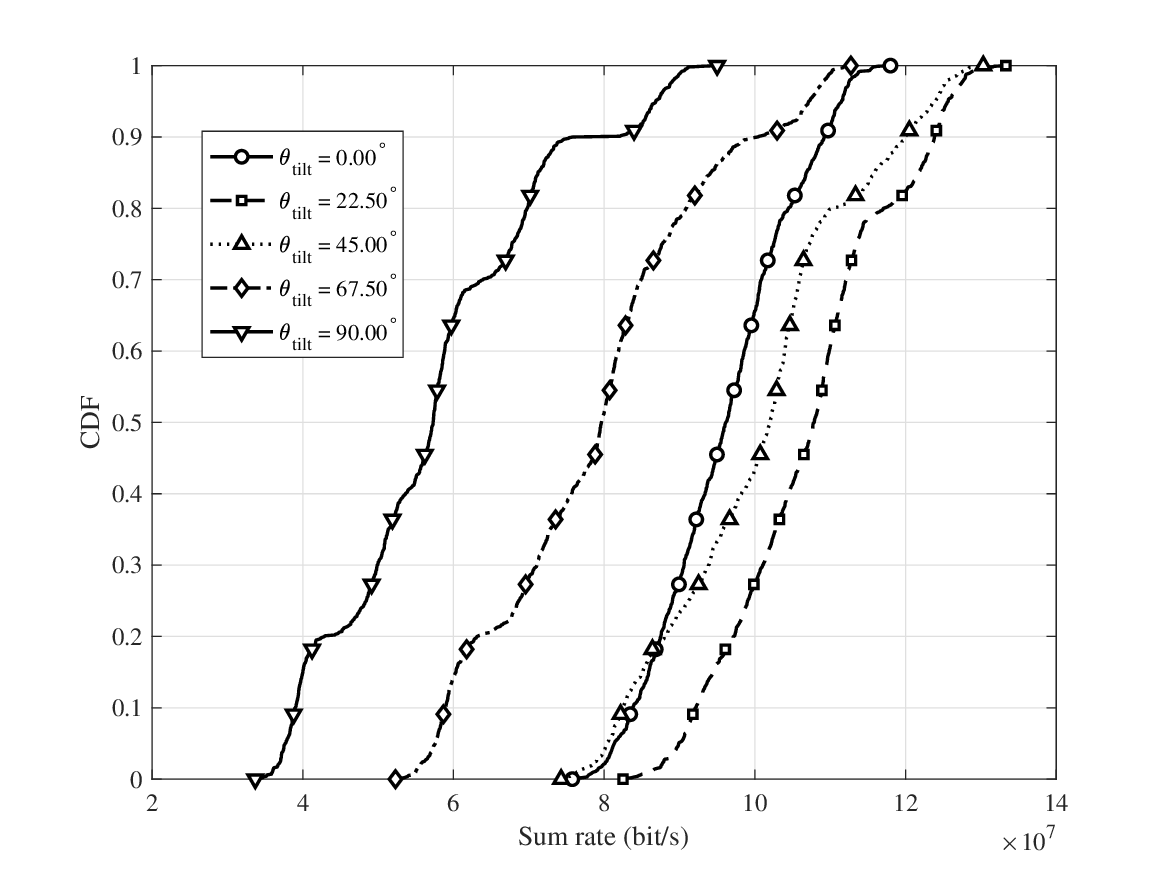}
\caption{Sum-rate CDFs of the truncated cone array for different tilt angles.}
\label{Fig:tilt_sum_rate}
\end{figure}
The proposed truncated cone array is illustrated in the fourth row of Table~\ref{tab:architecture_summary}.
In the cylindrical array, the side-element boresights are parallel to the ground, corresponding to an off-nadir angle of $90^\circ$. However, for a HAPS at an altitude of $20$ km with a coverage radius of $100$ km, ground-user directions span off-nadir angles from $0^\circ$ to approximately $78.7^\circ$. The truncated cone array tilts the side-element boresights downward to reduce this angular mismatch and improve antenna gain toward ground users. Meanwhile, these boresights are distributed around the full azimuth, allowing different subsets of elements to provide favorable gain toward users in different azimuth directions.
As shown in Fig.~\ref{Fig:sum_rate_R100}, the truncated cone array achieves a higher sum rate than the cylindrical array, highlighting the benefits of introducing a tilted side surface to better serve widely dispersed users.

To further examine the impact of the side-surface tilt, Fig.~\ref{Fig:tilt_sum_rate} presents the CDF of the uplink sum rate for tilt angles of $0^\circ$, $22.5^\circ$, $45^\circ$, $67.5^\circ$, and $90^\circ$. Among the tested angles, $22.5^\circ$ achieves the highest sum rate over most of the CDF range, although its curve crosses that of $45^\circ$ near the upper tail. Increasing the tilt angle beyond $45^\circ$ substantially degrades performance, with $90^\circ$ yielding the lowest sum rate. These results highlight the importance of tilt-angle selection and favor an intermediate tilt angle under the considered simulation setting.
From a practical perspective, the truncated cone’s regular rotational geometry supports sector-based subarray partitioning for hybrid beamforming, while its inclined surface requires tailored element mounting and feeding arrangements. It is therefore a promising option for wide-area coverage, provided that the chosen tilt and array dimensions remain compatible with the platform’s payload and aerodynamic constraints.

\subsubsection{Truncated Hemispherical Array}
The proposed truncated hemispherical array is illustrated in the last row of Table~\ref{tab:architecture_summary}.
Relatively few elements of the hemispherical array point toward
the nadir, limiting the antenna gain available
to users directly beneath the HAPS.
To strengthen antenna gain in this direction, the proposed architecture incorporates a downward-facing bottom panel, increasing the proportion of elements whose boresights align with nadir users. The remaining curved surface provides boresight diversity in both azimuth and elevation, allowing different subsets of elements to support users at different ground distances and azimuth angles.
As shown in Fig.~\ref{Fig:sum_rate_concentrated}, it achieves the second-highest sum rate among the considered antenna arrays, following the UPA, demonstrating the benefits of combining enhanced nadir coverage with diverse element orientations. From a practical perspective, the truncated hemispherical array retains the fabrication, feeding, calibration, and RF-chain mapping challenges of the hemispherical array. Integrating the bottom panel further complicates structural assembly.

\subsection{Discussion}
The performance differences among the considered architectures can be understood from four aspects: elevation-angle coverage, element-boresight distribution, spatial correlation, and channel separability. Planar arrays concentrate high gain near the nadir, while cylindrical arrays add horizontal side-element boresights to support larger off-nadir angles. The truncated cone and 3GPP hexagonal arrays tilt their side elements downward, reducing boresight mismatch with ground users. Hemispherical arrays distribute boresights across both azimuth and elevation, providing favorable gain toward users in different directions. The truncated hemispherical array further strengthens nadir gain through its bottom panel. 
\par
Array geometry and element orientations also affect spatial correlation and channel separability. Planar arrays rely primarily on geometry-dependent phase differences for user separation, whereas cylindrical and truncated cone arrays introduce additional gain-profile differences across azimuth directions. Hemispherical and truncated hemispherical arrays extend this diversity to elevation, allowing users at different ground distances to exhibit distinct gain profiles as well. Diverse element orientations and positions can also modify inter-element spatial correlation by shaping the gain and phase responses to multipath components. Together, these features can support greater effective channel dimensionality and improved user-channel separability, facilitating interference suppression.
\par

These characteristics help explain the observed performance advantages of the truncated cone and truncated hemispherical arrays. For users sparsely distributed over a wide area, the truncated cone array combines downward-tilted side elements with full azimuthal coverage, improving boresight alignment across the service area and achieving the best performance in the considered scenarios. 
For users densely concentrated beneath the HAPS, the planar array performs best because all elements face the nadir. The truncated hemispherical array achieves the next-best performance in Fig.~\ref{Fig:sum_rate_concentrated}, benefiting from the enhanced nadir gain provided by its bottom panel while retaining curved-surface orientation diversity in both azimuth and elevation to help  reduce spatial correlation and improve channel separability.
It also achieves performance comparable to the truncated cone array for widely distributed users, as shown in Fig.~\ref{Fig:sum_rate_R100}. 

\par

Nevertheless, these communication-performance advantages do not by themselves establish hardware efficiency or practical feasibility. The comparisons in this article characterize performance under the considered simulation assumptions, without quantitatively evaluating payload mass, power consumption, mechanical complexity, RF-chain requirements, or implementation cost. Practical architecture selection therefore requires joint consideration of the observed performance benefits and these implementation constraints, whose quantitative evaluation remains an important direction for future research.

\section{Open Research Challenges}
The antenna architectures discussed in this article illustrate
how array geometry and element orientation can improve coverage
and spatial multiplexing in HAPS networks. However, fixed geometries,
simplified antenna models, and substantial RF hardware demands
leave several challenges unresolved. These limitations motivate
further research into adaptive array configurations, electromagnetic-aware
design, and hardware-efficient implementations. The following
subsections outline promising directions for addressing these
challenges in future HAPS networks.

\subsection{Movable Antenna-Enabled HAPS Arrays}
Existing HAPS arrays use fixed geometries and element orientations, limiting their adaptability to changing user distributions and traffic demands. Movable antenna (MA) technology introduces additional flexibility by allowing antenna elements to adjust their positions \cite{zhu2023movable}. As shown in Figs.~\ref{Fig:sum_rate_R100} and ~\ref{Fig:sum_rate_concentrated}, downward-facing arrays achieve higher sum rates for users concentrated beneath the HAPS, whereas three-dimensional arrays better support widely distributed users. These results motivate adaptive array configurations. 
A concrete research problem is to jointly optimize element positions, tilt angles, and beamforming to maximize the sum rate or cell-edge SINR under movement, spacing, energy, and payload constraints. The coupling between geometry-dependent channel responses and beamforming makes this problem non-convex. A two-timescale approach could reduce the reconfiguration burden by adapting the geometry to long-term user distributions and updating beamforming using instantaneous channel state information (CSI).

\subsection{Electromagnetic–Communication Co-design}
The simplified element radiation patterns used in the preceding comparison do not fully capture mutual coupling, cross-polarization, or hardware impairments. This limitation motivates electromagnetic–communication co-design, which connects practical electromagnetic behavior with system-level performance.
Advances in pattern-reconfigurable antennas \cite{wang2026transformer}
and techniques for reducing mutual coupling motivate the joint design of antenna electromagnetic characteristics and beamforming for HAPS arrays.
A concrete research problem is to jointly optimize array geometry, element orientations, reconfigurable radiation patterns, and beamforming using models that capture mutual coupling and cross-polarization. Practical hardware impairments, such as phase noise, amplifier nonlinearity, and finite-resolution phase shifters, should also be incorporated to assess their impact on SINR and develop impairment-aware beamforming and calibration techniques.



\subsection{Non-uniform Antenna Arrays for HAPS}
The regular layouts and uniform spacing considered in the preceding comparison restrict the flexibility of element placement for controlling sidelobes and inter-user channel correlation. This motivates non-uniform array designs that adapt element placement to the array surface and coverage requirements.
Non-uniform layouts, such as spiral-based designs \cite{tuzi2023satellite}, offer additional flexibility to reduce spatial correlation and sidelobe levels while preserving wide-area coverage with fewer antenna elements, thereby improving spectral efficiency and reducing hardware cost.
A concrete research problem is to optimize element placement on a given three-dimensional surface to suppress sidelobes and improve user channel separability under element-count and minimum-spacing constraints. Since surface curvature and coverage requirements affect where antenna elements should be placed, sparse array optimization or learning-based methods can be used to select element locations
on the array surface to achieve the desired coverage with a limited number of elements.



\subsection{Metamaterial-Enabled HAPS Antenna Arrays}
The RF-chain and feeding-network complexity motivates alternative architectures for implementing large HAPS arrays with lower hardware and power requirements.
Metasurface-based antennas offer a potential approach by integrating tunable radiating elements with shared feeding structures to  tailor electromagnetic responses \cite{caloz2005electromagnetic}. 
Tunable implementations enable beam steering and radiation pattern reconfiguration, offering potential reductions in array size and hardware complexity for HAPS platforms with stringent payload and power constraints. 
A concrete research problem is to jointly optimize programmable element states and beamforming to maximize energy efficiency under coverage and power constraints. The design should account for discrete tuning states, electromagnetic coupling, and losses, while incorporating total hardware power consumption to assess the achievable energy savings.


\subsection{Artificial Intelligence (AI)-Driven HAPS Antenna Design}
The preceding research directions involve strongly coupled geometric, electromagnetic, and hardware variables, making repeated simulation and joint optimization computationally demanding. 
This motivates AI-assisted antenna design using surrogate modeling and data-driven optimization to accelerate performance evaluation and efficiently explore high-dimensional design spaces \cite{Akinsolu_antenna}.
A concrete problem is to train a surrogate model that maps the full set of design variables, including geometry, element placement, orientation, and RF chain and subarray configuration, directly to communication performance (e.g., energy efficiency). This can reduce the need for costly full-wave electromagnetic simulations and enable joint optimization of these coupled design variables. Such a framework could also incorporate attitude uncertainty as an input, allowing the model to learn robust designs under different perturbation scenarios rather than optimizing only for a fixed HAPS position.

\section{Conclusions}
As HAPS evolves toward a key component of beyond-6G wireless networks, antenna array architecture is becoming increasingly important for achieving reliable wide-area coverage and high-capacity communications. This paper has reviewed representative HAPS antenna array architectures and introduced two novel architectures to further enhance communication performance. Through a unified performance evaluation framework, the advantages and limitations of different antenna architectures have been systematically demonstrated, providing valuable insights into their design evolution. Finally, several open research challenges have been identified, highlighting promising directions for the development of intelligent, adaptive, and high-performance HAPS antenna arrays.

\bibliography{my_bibliography}
\bibliographystyle{IEEEtran}
\end{document}